\documentclass[aps,prd,10pt]{revtex4-1}
\usepackage{floatpag}
\usepackage{amsfonts}
\usepackage{amsmath}
\usepackage{amssymb}
\usepackage{latexsym}
\usepackage{times}
\usepackage{bm}
\usepackage{graphicx}

\begin{document}
\title{\huge Dynamical rotational frequency shift}
\author{\Large Iwo Bialynicki-Birula}\email{\large birula@cft.edu.pl}
\affiliation{\large Center for Theoretical Physics, Polish Academy of Sciences, Al. Lotnik\'ow 32/46, 02-668 Warsaw, Poland}
\author{\Large Zofia Bialynicka-Birula}
\affiliation{\large Institute of Physics, Polish Academy of Sciences, Al. Lotnik\'ow 32/46, 02-668 Warsaw, Poland}

\maketitle

\section{Introduction}

The term rotational frequency shift (RFS) has been used in different contexts and it was given different meanings \cite{ga}--\cite{n1}. Other terms have also been used (azimuthal Doppler shift, angular Doppler shift) to describe various related phenomena. In this article we stick to the meaning of the rotational frequency shift given by us in Ref.~\cite{bb1}. In order to make a clear distinction between our RFS and other related shifts we use the term dynamical RFS (DRFS). We will study the spectral properties of radiation emitted by rotating quantum sources.

Radiation emitted by sources in motion looks different when observed in the laboratory frame. Frequency shift can only be determined for monochromatic waves. In general, a monochromatic field will loose this property when observed from a moving frame. Therefore, to observe a frequency shift we have to restrict ourselves to some special forms of radiation, which fully preserve their monochromaticity when the frame of reference is changed.

When the source moves with constant velocity, one observes only the well known Doppler shift. In this case a special role is played by plane waves characterized by their wave vectors. We may invoke the relativistic transformation properties of the wave vector to derive the change of the frequency. As a result, all inertial observers see such waves as monochromatic plane waves with a shifted frequency and a transformed direction of propagation. This effect can also be deduced from the transformation laws of the photon energy-momentum four-vector $p_\mu=\hbar k_\mu$. For the uniform motion this kinematical Doppler shift is the only effect.  According to the principle of relativity there is no preferred inertial frame of reference. Therefore, the properties of the moving source are not affected by its uniform motion.

When the source rotates with respect to an inertial frame, the symmetry between the two frames is lost. The source will ``know'' that it is moving because it ``feels'' inertial forces. These forces affect the dynamics of the radiation process. In particular, the energy levels of a rotating source are modified. This is the first contribution to the dynamical rotational frequency shift. The second contribution is due to the transformation properties of the electromagnetic field. The radiation emitted by a rotating source looks different when observed in the laboratory frame because when the frame is changed, the electric and magnetic field vectors must be transformed accordingly. In the case of rotation, the role of the photon momentum is played by angular momentum. Monochromatic waves with a given projection of the angular momentum on the axis of rotation will remain monochromatic for all observers rotating around this axis. Their frequency will be shifted by $M\Omega$, where $M$ is a natural number that determines the projection of the angular momentum on rotation axis and $\Omega$ is the frequency of rotation. This kinematical shift of frequency is the counterpart of the Doppler shift and it has been predicted and observed in Refs.~\cite{ga}--\cite{n1}. Our aim here is to study an interplay between this purely kinematical shift and the shift caused by the rotation of the source. The combined effect is the dynamical rotation frequency shift (DRFS).

The dynamical rotational frequency shift in its pure form so far has not been seen. However, its close counterpart was detected in an experiment \cite{mhs} on thioformaldehide molecules that were forced to rotate by an applied circularly polarized microwave.

The study of the DRFS involves the energy levels of the emitting system. Therefore, it requires a quantum mechanical setting and we begin with a suitable form of description which will be applied first to the Doppler shift.

\section{Doppler shift}

In order to describe in the next section the DRFS shift in a fully quantum-mechanical setting, we shall introduce first the analogous formulation for the standard Doppler shift, even though in this simple case this description may look overly complicated. Let us imagine an atom moving with constant velocity ${\bm v}$. The radiation process in the laboratory frame is described by the Schr\"odinger equation,
\begin{align}\label{sch}
i\hbar\partial_t|\psi(t)\rangle=H(t)|\psi(t)\rangle,
\end{align}
with the following Hamiltonian (for definiteness we consider one electron as a source of radiation):
\begin{align}
H(t)&=H_A(t)+H_F+H_I,\label{ham1}\\
H_A(t)&=\frac{{\bm p}^2}{2m}+V({\bm r}(t)),\label{ham2}\\
H_F&=\tfrac{1}{2}\!\int\!d^3r:\left[\epsilon_0{\bm E}^2({\bm r})+{\bm B}^2({\bm r})/\mu_0\right]:=\hbar\sum_i\omega_ia_i^\dagger a_i,\label{ham3}\\
H_I&=-\frac{e{\bm p}\!\cdot\!{\bm A}({\bm r})}{m}+\frac{e^2{\bm A}^2({\bm r})}{2m},\label{ham4}
\end{align}
where ${\bm E}({\bm r}),\;{\bm B}({\bm r})$ and ${\bm A}({\bm r})$ are the operators of the electric field, magnetic field and the vector potential. The sum over the modes of the electromagnetic field may also include integration. The time dependence of the potential $V$ is due to a motion of its center, ${\bm r}(t)={\bm r}-{\bm R}(t)$. For a uniform motion ${\bm R}(t)={\bm v}t$. For future reference, we have split the Hamiltonian into the atomic part, the field part, and the interaction part. The coupling of the electron to the radiation field will be treated as a perturbation.

We will now eliminate the time dependence of $H$ by a transformation to the moving frame. To this end, we transform the state vectors  according to the following unitary transformation $U_V(t)$:
\begin{align}\label{unitaryt}
U_V(t)=e^{-i{\bm v}\cdot{\bm P}_T\,t/\hbar},\\
|\psi(t)\rangle=U_V(t)|\tilde{\psi}(t)\rangle,
\end{align}
where ${\bm P}_T$ is the operator of the total momentum of the system. This transformation removes the time dependence of the Hamiltonian since
\begin{align}\label{transt}
U_V^\dagger(t){\bm r}(t)U_V(t)=U_V^\dagger(t)\left({\bm r}-{\bm v}t\right)U_V(t)={\bm r},
\end{align}
but it also produces an additional term in the Schr\"odinger equation,
\begin{align}\label{scht}
i\hbar\partial_t|\tilde{\psi}(t)\rangle=\left[H(t)|_{t=0}-{\bm v}\!\cdot\!{\bm P}_T\right]|\tilde{\psi}(t)\rangle.
\end{align}
The additional term in the Hamiltonian comes from the time dependence of the transformation.

The atomic part ${\bm p}$ of the total momentum modifies the atomic Hamiltonian,
\begin{align}\label{hama}
\tilde{H}_A&=H_A(t)|_{t=0}-{\bm v}\!\cdot\!{\bm p}.
\end{align}
It is important to note that this modification does not lead to a change of the energy level separation. It can be eliminated (apart from an additive constant $m{\bm v}^2/2$) by an optional unitary transformation
\begin{align}\label{transt1}
|\tilde{\psi}(t)\rangle=e^{im{\bm v}\cdot{\bm r}/\hbar}|\phi(t)\rangle,
\end{align}
that restores the standard form, ${\bm p}^2/2m+V({\bm r})$, of the atomic Hamiltonian. Thus, the separations of the atomic energy levels in a uniformly moving frame are not modified---an expression of the relativity principle.

The interaction Hamiltonian is not modified by the transformation $U_V(t)$, since it commutes with the total momentum. However, it acquires an additional term $e{\bm v}\!\cdot\!{\bm A}({\bm r})$ after the transformation (\ref{transt1}). This only changes slightly the photon emission probability but not the photon spectrum.

The field part ${\bm P}$ of the total momentum modifies the field Hamiltonian,
\begin{align}\label{hamf}
\tilde{H}_F&=H_F-{\bm v}\!\cdot\!{\bm P}=\hbar\sum_\lambda\int\!d^3k\,(\omega({\bm k})-{\bm v}\!\cdot\!{\bm k})a_\lambda^\dagger({\bm k})a_\lambda({\bm k}).
\end{align}
We have chosen here plane waves as the appropriate modes of the electromagnetic field. The modification of the field Hamiltonian is responsible for the Doppler shift. This follows from the energy conservation in the moving frame where the Hamiltonian is time independent. The eigenvalue of the photon Hamiltonian must match the energy difference of the atomic system $\omega-{\bm v}\!\cdot\!{\bm k}=\Delta E/\hbar$. Therefore, the frequency of the photon $\omega$ in the laboratory frame undergoes the Doppler shift,
\begin{align}\label{doppler}
\omega=\Delta E/\hbar+{\bm v}\!\cdot\!{\bm k}.
\end{align}
The quadratic Doppler shift does not appear here, since we are using the nonrelativistic theory.

\section{Dynamical rotational frequency shift}

We shall now repeat the procedure that led to the standard Doppler effect, replacing translation by rotation in the time-dependent potential. In order to remove the time dependence of the Hamiltonian, we apply the unitary transformation $U_{\Omega}(t)$,
\begin{align}\label{unitaryr}
U_{\Omega}(t)=e^{-i{\bm\Omega}\cdot{\bm J}\,t/\hbar},
\end{align}
where ${\bm\Omega}$ is the angular velocity vector and ${\bm J}$ is the total angular momentum. This transformation is the counterpart of the transformation $U_V(t)$. It removes the time dependence of the potential since ${\bm\Omega}\!\cdot\!{\bm J}$ is the generator of rotations around the ${\bm\Omega}$ axis,
\begin{align}\label{transr}
U_{\Omega}^\dagger(t){\bm r}(t)U_{\Omega}(t)={\bm r},
\end{align}
The additional term in the Schr\"odinger equation,
\begin{align}\label{schr1}
i\hbar\partial_t|\tilde{\psi}(t)\rangle=\left[H(t)|_{t=0}-{\bm\Omega}\!\cdot\!{\bm J}\right]|\tilde{\psi}(t)\rangle,
\end{align}
is the sum of the atomic part ${\bm\Omega}\!\cdot\!{\bm L}$ and the field part ${\bm\Omega}\!\cdot\!{\bm M}$.

The Hamiltonian in the rotating frame cannot be identified with the energy operator. However, its eigenvalues---the quasi-energies---are time independent and they may be used to follow the energy transfers between the atom and the field. Any decrease in the quasi-energy of the atom results in the corresponding increase in the photon quasi-energy.

In order to write down the field Hamiltonian in an explicit form we choose the decomposition of the electromagnetic field into different field modes than in the previous section. We choose the axis of rotation along the $z$ axis and instead of the plane-wave modes we employ the cylindrical modes labeled by the frequency $\omega$, the projection $M$ of the total angular momentum on the $z$ direction, the $z$ component of the wave vector $k_z$ and the helicity $\chi$. With this choice, the field Hamiltonian after the unitary transformation is
\begin{align}\label{hamfr}
\tilde{H}_F&=\hbar\sum_{M,\chi}\int_0^\infty\!d\omega(\omega-\Omega M)\int_{-\infty}^\infty\!dk_z\, a^\dagger(\omega,M,k_z,\chi)a(\omega,M,k_z,\chi).
\end{align}
As in the case of the Doppler shift, we may say that in the rotating frame each photon acquires additional energy. In the present case the energy difference is equal to $\hbar\Omega M$. Again there is no change in the interaction Hamiltonian since it is invariant under rotations, $H_I$ commutes with the total angular momentum.

Despite formal similarities, there is a big difference between the translational and the rotational motion that affect the atomic Hamiltonian. The energy levels of this Hamiltonian are now modified and not just by a trivial constant term as was the case for the Doppler shift. We shall analyze these modification in two steps.

First, let us assume that the potential energy $V$ has cylindrical symmetry. In this case, the $\Omega$-dependent part of the atomic Hamiltonian $\tilde{H}_A$,
\begin{align}\label{hamr}
\tilde{H}_A&=H_A(t)|_{t=0}-{\bm\Omega}\!\cdot\!{\bm L}=H_A(t)|_{t=0}-\Omega(xp_y-yp_x).
\end{align}
commutes with the remaining part but it cannot be removed by a unitary transformation. The quasi-energy spectrum is modified and it resembles the atomic spectrum in a constant magnetic field. Each degenerate level characterized by the maximal value $m_{\rm max}$ of the $z$ component of angular momentum is split into $2m_{\rm max}+1$ equally spaced Zeeman-like sublevels separated by $\hbar\Omega$. This level splitting, however, for systems with cylindrical symmetry will never be seen. The explanation of this fact within our model involves a cancelation of two effects. The quasi-energy spectrum of the atomic electron is $E_q-\hbar\Omega m_z$, where $q$ stands for a set of quantum numbers characterizing different eigenstates of the Hamiltonian $H_A(t)|_{t=0}$. Owing to the quasi-energy conservation, the photon emitted in an atomic transition between two states with quantum numbers $(q,m_z)$ and $(q',m_z')$ has its quasi-energy $\hbar(\omega-\Omega M)$ equal to the difference $E_q-E_{q'}-\hbar\Omega(m_z-m_z')$. Thus, the photon frequency is given by the formula
\begin{align}
\omega=(E_q-E_{q'})/\hbar-\Omega(m_z-m_z')+\Omega M.
\end{align}
For systems with cylindrical symmetry two last terms cancel due to the conservation of angular momentum and we are left with the standard Bohr formula. This confirms a general rule that a rotation of a cylindrically symmetric {\em quantum object} around its symmetry axis is unobservable.

Next, we consider the potential $V({\bm r})$ that is not invariant under rotations. In this case, the angular momentum operator does not commute with the atomic Hamiltonian. Therefore, the change of the spectrum of $\tilde H_A$ induced by rotation does not have such a simple structure, as in the rotationally symmetric case. In order to find the spectrum of quasi-energies we must diagonalize the complete Hamiltonian and not separately its two commuting parts. The spectrum of quasi-energies, in general, is described by some nonlinear function ${\cal E}(q,m_z,\Omega)$ of $\Omega$. Therefore, the dependence on $\Omega$ will not be canceled by the frequency shift $\Omega M$ of the photon, as was the case for cylindrically symmetric systems. The resulting frequency of the observed photon
\begin{align}\label{rfs}
\omega(\Omega)={\cal E}(q,m_z,\Omega)-{\cal E}(q',m_z',\Omega))/\hbar+\Omega M
\end{align}
differs from the frequency of a photon emitted by an atom at rest. The difference $\omega(\Omega)-\omega(0)$ is the dynamical frequency shift of Ref.~\cite{bb1}. In the next section we illustrate these properties with the help of specific examples.

\section{Atom on a turntable}

The title of this section is not to be taken literally (turntables rotate too slowly) but it conveys the right meaning. We shall study the changes of the atomic spectrum due to the orbital motion of its center. In this case, an atomic electron is moving in the potential $V({\bm r}-{\bm R}(t))$, where ${\bm R}(t)$ traces a circle with constant velocity.

We shall start by considering first an exactly soluble case: the harmonic oscillator. The Hamiltonian (\ref{ham2}) in this case is
\begin{align}\label{fh}
\frac{{\bm p}^2}{2m}+\frac{m\omega_0^2}{2}\left({\bm r}-{\bm R}(t)\right)^2,
\end{align}
where $\omega_0$ characterizes the spring tension of the oscillator. Due to the invariance of the distance under rotations,
\begin{align}
|{\bm r}-{\bm R}(-t)|=|{\bm r}(t)-{\bm R}|,
\end{align}
we can transfer time dependence to the position of the electron. Time independent atomic Hamiltonian (\ref{hamr}) obtained after the unitary transformation (\ref{transr}) is
\begin{align}\label{fh1}
{\tilde H}_A=\frac{{\bm p}^2}{2m}+\frac{m\omega_0^2}{2}\left({\bm r}-{\bm R}\right)^2-{\bm\Omega}\!\cdot\!{\bm L}.
\end{align}
It is convenient to move the origin of the coordinate system from the rotation axis to the center of the potential. This is accomplished by the unitary transformation $U_S$ that simply shifts the origin of the coordinate system by ${\bm R}$,
\begin{align}\label{unitr}
U_S=\exp[-i{\bm R}\!\cdot\!{\bm P}_T/\hbar].
\end{align}
After this transformation the three parts of the Hamiltonian of the system take on the form
\begin{align}
U_S^\dagger{\tilde H}_AU_S&=\frac{{\bm p}^2}{2m}+\frac{m\omega_0^2}{2}{\bm r}^2-{\bm\Omega}\!\cdot\!{\bm L}-{\bm v_c}\!\cdot\!{\bm p},\label{sha}\\
U_S^\dagger{\tilde H}_IU_S&=-\frac{e{\bm p}\!\cdot\!{\bm A}({\bm r})}{m}+\frac{e^2{\bm A}^2({\bm r})}{2m},\label{shi}\\
U_S^\dagger{\tilde H}_FU_S&=H_F-{\bm\Omega}\!\cdot\!{\bm M}-{\bm v_c}\!\cdot\!{\bm P},\label{shf}
\end{align}
where ${\bm v_c}$ is the orbital velocity of the center of the potential,
\begin{align}\label{velo}
{\bm v_c}={\bm\Omega}\times{\bm R}.
\end{align}
Additional terms proportional to ${\bm v_c}$ in the atomic and the field Hamiltonians arise because the total momentum ${\bm P}_T$ does not commute with angular momentum: from the commutation relations between momentum and angular momentum we obtain for the atomic part
\begin{align}\label{add}
U_S^\dagger{\bm\Omega}\!\cdot\!{\bm L}U_S={\bm\Omega}\!\cdot\!{\bm L}-i\hbar[{\bm R}\!\cdot\!{\bm p},{\bm\Omega}\!\cdot\!{\bm L}]={\bm v_c}\!\cdot\!{\bm p}.
\end{align}
An analogous relation holds for the field part. The atomic and field Hamiltonians (\ref{sha}) and (\ref{shf}) reduce to the Hamiltonians (\ref{hama}) and (\ref{hamf}) in the limit, when $\Omega\to 0$ and $R\to\infty$, while the orbital velocity ${\bm v_c}$ is kept fixed. It must be so, because in this limit, as in the case of the Doppler shift, the motion of the atom takes place along a straight line with constant velocity. In fact, Doppler broadenings in the spectra of rotating stars rely on this limiting case.

The atomic Hamiltonian can be simplified by an additional unitary transformation $U$ that removes the last term in (\ref{sha}) and reduces the problem to the standard case of a harmonic oscillator.
\begin{align}\label{unit}
U=\exp[i{\bm a}\!\cdot\!{\bm r}/\hbar-i{\bm b}\!\cdot\!{\bm P}_T/\hbar].
\end{align}
When the vectors ${\bm a}$ and ${\bm b}$ are chosen as
\begin{align}\label{ab}
{\bm a}=\frac{m\omega_0^2}{\omega_0^2-\Omega^2}{\bm\Omega}\times{\bm R},\quad
{\bm b}=\frac{\Omega^2}{\omega_0^2-\Omega^2}{\bm R},
\end{align}
the transformed Hamiltonian ${\tilde H}_A$ takes on the standard, rotationally symmetric form,
\begin{align}\label{fh3}
{\hat H}_A=U^\dagger{\tilde H}_AU=\frac{{\bm p}^2}{2m}+\frac{m\omega_0^2}{2}{\bm r}^2-{\bm\Omega}\!\cdot\!{\bm L}-\frac{m\omega_0^2\Omega^2R^2}{2(\omega_0^2-\Omega^2)}.
\end{align}
The field Hamiltonian is not modified by the transformation (\ref{unit}) but the interaction Hamiltonian changes as follows:
\begin{align}\label{fh4}
{\hat H}_F&=U^\dagger{\tilde H}_FU={\tilde H}_F,\\
{\hat H}_I&=U^\dagger{\tilde H}_IU=-\frac{e{\bm p}\!\cdot\!{\bm A}({\bm r})}{m}+\frac{e^2{\bm A}^2({\bm r})}{2m}-\frac{\omega_0^2}{\omega_0^2-\Omega^2}e{\bm v}_c\!\cdot\!{\bm A}({\bm r}).
\end{align}
After the unitary transformations the atomic and field Hamiltonians are the same as for a harmonic oscillator placed on the rotation axis. If it were not for the last term in the interaction Hamiltonian, this system would have exact rotational symmetry and no effect of rotation would be seen. This additional term does not change the spectrum but it merely changes the photon transition probabilities. The absence of any changes in the spectrum is exceptional and it is specific to the harmonic potential.

We shall consider now a more realistic case of an atom on a turntable by replacing the harmonic potential by the Coulomb potential. This will lead to significant changes in the spectrum. Following the procedure applied in the previous case, after the first few steps, we obtain the following Hamiltonian:
\begin{align}
U_S^\dagger{\tilde H}_AU_S&=\frac{{\bm p}^2}{2m}-\frac{\gamma}{|{\bm r}|}-{\bm\Omega}\!\cdot\!{\bm L}-{\bm v_c}\!\cdot\!{\bm p},\label{sca}
\end{align}
where $\gamma=Ze^2/4\pi\epsilon_0$.
The interaction and the field Hamiltonians are given by the same formulas (\ref{shi}) and (\ref{shf}). The quasi-energy spectrum of this Hamiltonian cannot be determined exactly. However, the following unitary transformation:
\begin{align}
U=\exp\left[im{\bm r}\!\cdot\!{\bm v}_c/\hbar\right]
\end{align}
produces a familiar expression
\begin{align}\label{fc}
U^\dagger\tilde{H}_AU=\frac{{\bm p}^2}{2m}-\frac{\gamma}{|{\bm r}|}-e{\bm{\mathfrak E}}\!\cdot\!{\bm r}-\frac{e}{2m}{{\bm{\mathfrak B}}}\!\cdot\!{\bm L}-\frac{mR^2\Omega^2}{2},
\end{align}
where we introduced fictitious electric and magnetic fields that mimic the action of inertial forces
\begin{align}\label{mim}
e{\bm{\mathfrak E}}=m\Omega^2{\bm R},\quad e{\bm{\mathfrak B}}=2m{\bm\Omega}.
\end{align}
In this way, we reduced our task to the thoroughly studied problem of an electron moving in the Coulomb potential in the presence of additional crossed fields: the electric field ${\bm{\mathfrak E}}$ and the magnetic field ${\bm{\mathfrak B}}$. The quadratic in  ${\bm{\mathfrak B}}$ (diamagnetic) term is absent here which means that we may use the analogy between the rotation and the magnetic field only for relatively weak fields. The additive constant (last term) shifts the whole spectrum but it does not change the level separation.

The spectrum of the Hamiltonian (\ref{fc}) in the weak-field regime has been found already in old quantum theory \cite{ep}. In quantum mechanics, the solution of this problem is an example of perturbation theory with degenerate unperturbed spectrum. The energy levels of the Hamiltonian (\ref{fc}) in the lowest order in ${\bm{\mathfrak E}}$ and ${\bm{\mathfrak B}}$ are (cf., for example, \cite{pdg})
\begin{align}\label{sp}
{\cal E}(n,m_z,\Omega)=-m\left(\frac{Ze^2}{4\pi\epsilon_0\hbar}\right)^2\frac{1}{2n^2}
-\hbar m_z\sqrt{\left(e{\bm{\mathfrak B}}/2m\right)^2+\left(\frac{12\,\pi\epsilon_0\hbar}{2Zem}\right)^2n^2{\bm{\mathfrak E}}^2}.
\end{align}
Thus, the inertial forces in a rotating frame lead to modifications of the energy levels similar to the dynamical Stark shift caused by an electromagnetic wave. The quantum number $m_z=-m_{\rm max},\dots,m_{\rm max}$ plays the same role as in the rotationally symmetric case.

Upon inserting the expressions (\ref{mim}) into the formula (\ref{sp}), we obtain the following atomic spectrum modified by rotation:
\begin{align}\label{sp1}
{\cal E}(n,m_z,\Omega)=-\frac{mv_a^2}{2n^2}
-\hbar\Omega m_z\sqrt{1+\left(\frac{3nR\Omega}{2v_a}\right)^2},
\end{align}
where $v_a$ is the characteristic velocity of the electron in an atom,
\begin{align}\label{vel}
v_a=\frac{Ze^2}{4\pi\epsilon_0\hbar}.
\end{align}
For $Z=1$ this velocity is equal to the speed of light multiplied by the fine-structure constant, $v_a=\alpha c\approx 2.2\cdot 10^6{\rm m/s}$.

The DRFS of a photon emitted by an orbiting atom in a transition $(n,m_z)\to(n',m_z')$, observed in the laboratory frame, according to the formula (\ref{rfs}), is
\begin{align}\label{drfs}
\omega(\Omega)-\omega(0)&=
-\Omega m_z\sqrt{1+\left(\frac{3nR\Omega}{2v_a}\right)^2}-\Omega m_z'\sqrt{1+\left(\frac{3n'R\Omega}{2v_a}\right)^2}+\Omega M\nonumber\\
&\approx\frac{9\pi^2(n'^2 m_z'-n^2 m_z)}{2}\Omega\left(\frac{v_c}{v_a}\right)^2,
\end{align}
The smallness of the velocity $v_c$ of the potential center as compared to $v_a$ justifies the approximation of the square root by the first two terms of the Taylor expansion. The linear dependence of the DRFS on the quantum numbers $m_z$ that label the projections of the angular momentum on the axis of rotation means that the observed radiation can be red-shifted or blue-shifted. These quantum numbers are the analogues of the projections of the photon momentum on the direction of motion in the standard Doppler shift.

In the case of an orbiting atom one may also consider the linear Doppler shift due to the orbital velocity $v_c$. However, a direct comparison of the DRFS and the Doppler shift is not possible for two reasons. First of all, the orbital velocity is continuously changing its direction. In order to measure the standard Doppler shift of an orbiting atom, we must know its position on the orbit at the moment of the photon emission. This information is not available (except in the limiting case of $\Omega\to 0$, as described above) since the description of states according to the their quasi-energy is complementary (owing to the time-energy uncertainty relation) to the description that involves time. Second, these two effect rely on complementary descriptions of the radiation process. Photon eigenstates of $M_z$ contain the whole range of wave vectors and vice versa the eigenstates of momentum contain the whole range of angular momenta.

Even though we cannot compare directly the linear Doppler shift with the DRFS, we can do it for the quadratic Doppler shift. The quadratic Doppler shift in the direction perpendicular to the orbital plane and far from this plane is not sensitive to the position of the atom on its orbit. Therefore, the observation of the quadratic (transverse) Doppler shift on the axis of rotation $\omega(v/c)^2/2$ can be directly compared with the DRFS. Their ratio depends only on the ratio of two frequencies and not on the orbital velocity. Within the approximation employed in the formula (\ref{drfs}) it can be estimated as
\begin{align}\label{ratio}
9\pi^2(n'^2 m_z'-n^2 m_z)\frac{1}{\alpha^2}\frac{\Omega}{\omega},
\end{align}
where $\alpha$ is fine structure constant. This ratio may vary over a very wide range, from $10^{-10}$ for experiments using M\"ossbauer effect \cite{k,kymr} to as much as $10^3$ for rotating molecules emitting infrared photons.

\section{Observation of the dynamical rotational frequency shift}

The DRFS would be very hard to observe when the atom is placed on a macroscopic mechanical device. The fastest rotating devices, an ultracentrifuge, a gas turbine or a turbocharger, may reach at most a few thousand revolutions per second. For $R$ of the order of 0.1 meter this translates into the linear velocity of the order of 1000 m/s. Thus, the value of $(v/v_a)^2$, that appears in the formula (\ref{drfs}) is of the order of $10^{-6}$.

Can we make the measurement of the DRFS feasible, by resorting to microscopic rotating devices? Nature provided us with such objects in the form of various molecules. Molecules rotate very fast. Unfortunately, we also have to replace a macroscopic value of $R$ by a molecular bond length and this gives {\em the same effect} as a macroscopic ``turntable''. Indeed, with the values $\Omega\approx10^{13}/{\rm s}$ and $R\approx10^{-10}{\rm m}$ we again obtain $(v/v_a)^2\approx10^{-6}$. A large prefactor in (\ref{drfs}) increases the shift only by a factor of about 100 to the value of about $10^{-4}$ of the rotational frequency $\Omega$. Thus, in both cases the DRFS is very small and to detect it would require great ingenuity and determination. What has then been measured in the experiment described in Ref.~\cite{mhs}? To answer this question we have to extend our model to include an external electromagnetic field.

Molecules in gaseous phase at a given rotational temperature rotate with a random distribution of directions and a thermal distribution of absolute values of ${\bm\Omega}$. In order to organize this motion so that the signal would become detectable, the authors of the experiment \cite{mhs} used a circularly polarized microwave. The rotating electric field of the wave couples to the large electric dipoles of the thioformaldehyde molecules and makes them rotate in unison. To understand what frequency shift has been measured in this experiment, we shall take into account the role of a rotating electric filed in our turntable model. Such a field rotating with frequency $\Omega$ is represented by a dipole-coupling term, leading to the following form of the potential:
\begin{align}\label{pot1}
V({\bm r}(t))=-\frac{\gamma}{|{\bm r}-{\bm R}(t)|}-e{\bm r}\cdot{\bm E}(t).
\end{align}
In the nonrelativistic regime, the influence of the magnetic field of the wave can be neglected. After the transformation (\ref{transr}) the rotating electric field becomes frozen. In the case when our atom is embedded in a molecule, the electric field will be aligned with the molecular dipole. Thus, the time-independent Hamiltonian (\ref{fc}) in the presence of a rotating electric field has the form:
\begin{align}\label{f1}
H_{ff}=\frac{{\bm p}^2}{2m}-\frac{\gamma}{|{\bm r}|}-e\left({\bm{\mathfrak E}}+{\bm E}\right)\!\cdot\!{\bm r}-\frac{e}{m}{{\bm{\mathfrak B}}}\!\cdot\!{\bm L}-\frac{mR^2\Omega^2}{2}-e{\bm E}\!\cdot\!{\bm R}.
\end{align}
In general, the vectors ${\bm{\mathfrak E}}$ and ${\bm E}$ may have different directions. In the special case of a linear molecule they are collinear. Depending on whether they point in the same or in opposite directions, the effect will be enhanced or reduced. The two terms might even cancel each other. The spectrum of quasi-energies in the presence of a circularly polarized plane wave is still described by the formula (\ref{sp}) but one has to add to the pseudo-electric field ${\bm{\mathfrak E}}$ the true electric field ${\bm E}$. In this way we obtain for the quasi-energy spectrum
\begin{align}\label{sp3}
{\cal E}(n,m_z,\Omega)=-\frac{mv_a^2}{2n^2}
+\hbar\Omega m_z\sqrt{1+\left(\frac{3n(m\Omega^2{\bm R}+{e\bm E})}{2mv_a\Omega}\right)^2}.
\end{align}
Under the square-root sign in this expression there appears the sum of two forces: the centrifugal force $m\Omega^2{\bm R}$ and the electric force $e{\bm E}$. Their ratio for the values of the parameters of the experiment described in \cite{mhs} ($E=30{\rm kV/m},\;\Omega/2\pi=80{\rm MHz}$),is:
\begin{align}\label{est}
\frac{m\Omega^2R}{eE}=2.24\cdot10^{-10}\frac{(\Omega/2\pi[{\rm Hz}])^2R[{\rm m}]}{E[{\rm V/m}]}\approx 2.4\cdot10^{-9},
\end{align}
where the value of $R$ was chosen as equal to the atomic unit (Bohr radius) $5\cdot10^{-11}{\rm m}$. Thus, in this experiment the DRFS is nine orders of magnitude smaller than the dynamical Stark shift caused by the rotating electric field. It seems that in any experiment where a rotating electric field is used to control rotation, the influence of the electric field will completely overshadow the DRFS.

Of course, the turntable model does not represent correctly rotating molecules of thioformaldehyde in the experiment but it enables one to classify various terms in the formula for the DRFS. There are three contributions under the square-root sign in (\ref{sp3}) that differ in their dependence on the frequency of rotation $\Omega$: quadratic, quartic, and independent. It is the quartic term that describes the effect of quasi-energy level modifications in our model cause by the centrifugal force. Such term (or any other expression) was absent in the equations (7) and (8) of \cite{mhs}. Obviously, this omission was fully justified, as seen from our order of magnitude estimate.

\section{Conclusions}

We identified two sources of the dynamical rotational frequency shifts. The first part $\Omega M$ comes from the transformation properties the electromagnetic radiation under the change from the rotating frame to the inertial frame. The second part is due to the inertial forces acting on the emitter (or the absorber). For cylindrically symmetric emitters these two effect cancel each other. For asymmetric emitters the DRFS is in most cases very small but it might be observable in laboratory experiments.

\section*{Acknowledgements}

We acknowledge support from the Polish Ministry of Science and Higher Education under a grant for the years 2010-2012.

\end{document}